\documentclass[runningheads]{llncs}
\usepackage[T1]{fontenc}
\usepackage{graphicx}
\begin{document}
\title{Beyond Objects}
%
%
\author{Daniel Jackson\orcidID{0000-0003-4864-078X}} %
\authorrunning{D. Jackson}
\institute{MIT, Cambridge, MA\\
\email{dnj@mit.edu}\\
\url{http://people.csail.mit.edu/dnj}}

\maketitle              
\begin{abstract}

A core principle of object orientation---that the functionality of a system can be partitioned amongst objects that correspond to individuals in the problem domain---has influenced how software has been specified, designed and implemented for more than fifty years. Later developments in software engineering sought to build on this principle. But in fact this partitioning is neither natural nor straightforward, and the problems that these later developments sought to mitigate---the fragmentation and conflation of functionality---were often, in fact, the inevitable consequences of this founding principle. An easier path to addressing these problems therefore starts by going back, abandoning object orientation, and replacing it with an alternative approach that decouples the individuals of the problem domain from the modules that partition functionality. 

\end{abstract}

\noindent {\em To appear, Gedenkenschrift for Jean-Raymond Abrial, Springer Lecture Notes in Computer Science, Volume 16750.}

\section{Historical Preamble}

In 1974, Jean-Raymond Abrial presented a paper entitled \textit{Data Semantics} at an IFIP Working Conference on Data Base Management \cite{abrial1974datasemantics}. That paper came to be viewed as a seminal contribution to the theory of object orientation. Simula 67 \cite{dahl1966simula}, which was almost a decade old at that point, had introduced the key ideas of object-oriented programming, but only as an execution mechanism, and there was little understanding of objects as semantic entities independent of their implementation. 

Abrial's paper offered an abstract model of objects connected by relations, forming a bridge between the nascent world of object orientation and the relational data model that Codd had introduced a few years earlier \cite{codd1970relational}. Chen's entity relationship model would build on Abrial's model, and become the standard semantic view of relational data \cite{chen1976er}.

In Abrial's model, the state of a system is a collection of relations over object identities. These relations were global; there was no scoping or module structure. From the perspective of later developments in object orientation, this made Abrial's contribution suspect. Why did he not adopt the compelling object structuring that was already available in languages like Simula? 

Abrial seemed to realize that object-oriented structuring was not the right path to a more modular definition of behavior and state. In his subsequent work on Z \cite{abrial1980specification} (and later B \cite{abrial1996bbook}), he continued to view state relationally. Both Z and B offered structuring mechanisms for large specifications, with Z being particularly flexible in supporting a separation of concerns (exploited in work on view-based specification \cite{jackson1995structuring,ainsworth1994viewpoint,derrick1995crossview,finkelstein1992viewpoints}, which anticipates some of the ideas of this paper). B was more traditional than Z in this respect, but a proposal \cite{silva2010eventb} for adding event synchronization \cite{hoare1985csp} to Event-B \cite{abrial2010eventb} results in a language very similar to that described here.

Surprisingly, the researchers who developed modularity mechanisms for Z and B seem to have had rather limited goals---primarily easing the tasks of specification and verification. Papers and books about formal methods tend not to be concerned about the modularity of software, except insofar as it impacts the difficulty of constructing a one-time proof of correctness.

This seems to be a missed opportunity. In this paper, some classical ideas of abstract modeling from the field of formal methods---in particular relational states, events as state predicates, and event synchronization---are used to organize software in a new way that aligns with recent trends in industrial software development but offers improved separation of concerns and a more direct correspondence between problem domain and computational elements.

Seeds of these ideas can be found in early formal methods work. A 1987 specification of the CAVIAR reservation system by Bill Flinn and Ib Holm Sorensen that was included in Ian Hayes's case study collection \cite{hayes2001specification} proposed a parameterized module feature for Z, which it illustrated with a generic reservation module that could be applied (in different variants) to different kinds of resources (such as hotel rooms, conference rooms, and minibus seats). This specification inspired later work by Ian Hayes and Luke Wildman on modularity mechanisms for specification libraries \cite{hayes1992librariesz}. As the CAVIAR acknowledgments explain, these modularity ideas came from the original 1981 version of the specification, whose author was none other than Jean-Raymond Abrial himself. 

\section{Introduction}

On September 30, 1991, the cover of Business Week \cite{businessweek1991software} depicted a cartoon baby happily clicking away on a computer keyboard under the caption ``Software Made Simple.'' The article inside explained:

\begin{quote}
Instead of microchips, the software revolution will be built on so-called objects---simple, self-contained, reliable software components.
\end{quote}

Perhaps a revolution was indeed built on objects. Whether they provided reliability is hard to judge. But simple and self-contained? That objects would turn out not to be---and this paper will explain why.

In the conventional reading of the history of software engineering, object-oriented programming is one of its major successes. Following its invention more than 50 years ago, it has been widely applied in many successful systems. The technical challenges of designing and implementing object-oriented languages have been largely solved, and stable programming practices are reflected in well known patterns.

At the same time, the fundamental problems of software development have not gone away. Code is still hard to write and hard to read (and in many respects much harder than it was 30 years ago), and making the smallest changes to a large software system is still extremely challenging. While paying lip service to modularity, developers continue to entangle their code with couplings, with the popularity of the term `dependency hell' attesting to the pain this causes.

The advent of LLM-based coding promises to reduce dramatically the amount of human effort involved in programming. But code generated by LLMs follows the same practices that got human coders into trouble, and the use of LLMs seems poised to amass vast piles of technical debt---code that is fragile, unpredictable and near unmaintainable.

A brighter future, for human and LLM developers alike, is possible if we are prepared to reconsider the paradigms and assumptions that led us to this point. Many researchers and practitioners assume that object orientation was a major step forward, and that we simply need to refine and augment it with further advances. But many of the problems that currently plague software development are the direct and inevitable consequences of object orientation itself. The seeds of our contemporary problems were planted so long ago that a careful analysis will be needed to identify them and to trace their influence.

Object orientation is not a single idea, but a complex of ideas that have merged together in the popular conception of object-oriented programming. By disentangling these distinct notions and motivations that have become conflated over the years, we will be able to understand object orientation better and expose opportunities to use a subset rather than being committed to the entire package.

One particular principle, which brings two influential ideas together, has been the cause of many of our problems. It can be dropped without abandoning object-oriented programming in its entirety. But because this principle guides the high-level organization of most software systems today, it will need to be replaced by a different principle that achieves the intended goals without the liabilities.

\section{Object orientation: a complex of ideas}

The ideas of object orientation emerged in different communities that had different premises and motivations. A full accounting of these ideas would require a major study; here we will only sketch some concrete manifestations and allude to the intellectual trends that they represent.

Until the 1960s, software systems were elaborate scripts, usually executed in batch, that processed complex data often fed in on magnetic tape. The programs were organized into trees of functions that also maintained data structures in memory. The need to manage the complexity of these programs led in academia to the  notion of structured programming \cite{dijkstra1972structured} and in industry to structured analysis and design \cite{yourdon1989modern}, neither of which disrupted this view of a program as a tree of procedures.

\textbf{Tiny machines}. A very different paradigm emerged from the world of simulation, and viewed a software system as a collection of tiny machines or agents, each with its own state, communicating by sending messages to each other. This idea, often explained with a biological metaphor, is the basis of Simula 67, the first object-oriented language \cite{dahl1966simula}, Carl Hewitt's actor model \cite{hewitt1973actor}, and Alan Kay's conception of objects \cite{kay1993earlyhistory}. For Kay in particular, the idea of `late binding' was essential. This meant that the designer of an object could not rely on any assumptions that the recipient of a message would interpret it in any particular way. Message sending was thus a form of \textit{communication} and not \textit{delegation}: to ensure that the system as a whole achieved its intended goal, one would need to reason about the interactions between objects. Dijkstra and Hoare, in contrast, viewed a call from one function to another as a delegation of responsibility, so that the correctness of the caller could be established by assuming the correctness of the callee.

\textbf{The WIMP interface}. The ease of use of the WIMP (windows, icons, menus, pointer) interface, invented at Xerox PARC in the 1970s, came in large part from not having to recall the names of commands (or type them!). Instead, one only had to recognize the desired command in a list that popped up when you clicked on an icon. Crucially, the list was \textit{contextual}, showing all (and only) the relevant commands, being tailored to the kind of icon selected and the current state. In Smalltalk implementations, these icons were literally program objects, but the correspondence is hard to maintain. Deleting or renaming a file in a Unix-like file system, for example, requires mutating not the file itself but instead a directory entry associated with the folder of the selected icon.

\textbf{Abstract data types}. The idea of data abstraction emerged from two distinct threads. In 1972, David Parnas published his influential paper on modularity arguing that a decomposition based on encapsulated data structures was superior to one based on functional steps \cite{parnas1972criteria}. Around the same time, Barbara Liskov and Stephen Zilles proposed that data might be viewed purely in terms of its operations, without any commitment to a representation \cite{liskov1974adt}. Later, Liskov explained data abstraction as a way of overcoming the dichotomy between the built-in types of a programming language (such as integers and strings, which were accessible only through their operations) and user-defined types (which until then were expressed in a completely different fashion, with exposed data structures) \cite{liskov1977clu}. These ideas were later embodied in terms such as \textit{representation independence} (that it should be possible to change the representation of an abstract type without affecting its clients) and as \textit{information hiding} (although Parnas, who coined the term, meant it to refer to the hiding of all design decisions, not just those regarding representation). Amongst practitioners of object-oriented programming, the term \textit{responsibility-driven design} came to be used for this approach, although, oddly, the paper \cite{wirfsbrock1990rdd} that coined this term conflated data abstraction and `data-driven design' and seems to have been unaware that the approach it proposed had been standard in the data abstraction community for several years \cite{guttag1986abstraction}.

\textbf{Domain modeling}. An alternative to viewing a program as a datastore-processing script or as a network of communicating agents was instead to place a model of the problem domain at its center, and to regard the program as comprising (a) functions to update the model in reaction to changes from the outside world, and (b) functions to query the model or react (via triggers) to those changes. In Michael Jackson's JSD, perhaps the earliest example of this idea, the domain was modeled as a collection of entities, each with a set of instances and its own `life history' characterized by the allowable event sequences that it supported \cite{jackson1983systemdevelopment}. In this respect, JSD might have been inspired in part by Hoare's communicating sequential processes \cite{hoare1985csp},  which itself might have been influenced \cite{hoare2009mjdt} by Jackson's earlier work on JSP \cite{jackson1975principles}. Because JSP based the structure of a program on the structure of the data streams that it processed, both JSP and JSD were mistakenly classified as `data-driven approaches'. In fact, JSD is better seen as a precursor to the domain-centric object-oriented approaches such as OMT \cite{rumbaugh1991omt}, and to domain-driven design \cite{evans2003ddd} (especially in the practice of `event storming' in which the construction of the domain model begins, just as in JSD, with the listing of events). Of course the idea that a program might hold an explicit model of the world was already well established in systems that were built on databases, and in expert systems (which emerged in the 1960s). What distinguishes JSD and its successors, and that became normative in object-oriented systems, is the claim that \textit{all} software systems, even real-time control systems, can be treated profitably in this way, not just a specialized class of data processing systems. The approaches known as \textit{domain analysis} \cite{neighbors1980components} and \textit{domain engineering} \cite{bjorner2005software1} overlapped with domain modeling, but were motivated by something different: not so much the idea of maintaining a correspondence between the real world and a model inside the software, but rather the idea of exploiting commonalities across different systems that worked within the same application domain (such as railway signalling).

\textbf{Object identity}. Domain modeling only makes sense if the entities inside the software can be consistently mapped to their counterparts in the world. This means that each entity (instance) must have an identity that persists over time. This notion of object identity in programming is articulated in Abrial's seminal paper on data semantics \cite{abrial1974datasemantics}, which provides mechanisms for generating fresh objects with unique, persistent identities and binding them to variables. (If an externally visible identifier was required, it could be included as an attribute associated with an object.) This stands in marked contrast to relational schemas \cite{codd1970relational}, which had no such notion, and made it axiomatic that a relational table could not hold two tuples that contained the same values (modeling distinct objects with the same attributes). In this respect, Abrial provides the foundation for Chen's entity-relationship model \cite{chen1976er}, which reorganizes the relational model around object identities. The theory of normal forms in database schemas \cite{codd1972normalization} while elegant and powerful was in some respects a compensation for the lack of object identity; a schema based on entity-relationships is naturally in third normal form.

The notion of object identity does \textit{not} require that objects have internal structure; in Abrial's formalism, objects are associated with their attributes by global relations, and not by any kind of containment. The much later distinction between immutable objects (that have only value) and mutable objects (that have identity) is connected to this more fundamental notion of object identity only because, in an object-oriented implementation, the objects that correspond to real-world objects tend to be mutable.

\textbf{Objects as records}. The idea that an object can be represented as a record with a set of fields (or slots) defining its properties, and an independent identity (so that one can create two distinct records with the same field values) is now so familiar that we forget that it was at one time an innovation. A further enhancement, which had been available from the earliest days of LISP,  was to allow the values stored in the slots to include functions. Then came what William Cook argued \cite{cook1990oopvsadt} is the defining move for object-orientation: viewing the object as a record containing \textit{only} functions that act on other slots that are hidden from the object's users. These ideas emerged in the 1960s, first in Ivan Sutherland's Sketchpad \cite{sutherland1963sketchpad} and later in Simula 67 \cite{dahl1966simula} and then Smalltalk \cite{kay1993earlyhistory}.
 
\textbf{Classes and inheritance}. Neither classes nor inheritance are needed to support the principal mechanism of object-oriented languages (namely object-based method dispatch) but these features have been regarded by many as essential to object orientation. Class-based languages have dominated, although JavaScript, the most widely used language for web applications, only simulates classes with a syntactic sugar that hides its true prototype-based structure. Inheritance can help factor out shared code especially for graphical user interfaces, but has largely gone out of fashion in favor of delegation, due to the violations of modularity that it encourages \cite{liskov1994behavioral} and the inflexibility of single inheritance.

\section{The Roots of Trouble: Combining Two Ideas}

Debating what particular combination of these various ideas defines object orientation is not a productive enterprise. But two ideas have characterized much of the practice of object oriented development, and have had such a pervasive influence that they have shaped how software is constructed even when its developers would not describe their code as object oriented.

Either of these ideas alone might not have been problematic. In combination, however, they lead to a style of organization that tends to make software harder to build, understand and maintain.

The first idea is that desired system behaviors are obtained by assigning computations to objects. This follows from the notion of objects as tiny machines. Bertrand Meyer \cite{meyer1988oosc} calls this the \textit{Single Target} principle, and defines it like this: ``Every operation of object-oriented computation is relative to a certain object, the current instance at the time of the operation’s execution.''

The second idea is that computation is described with objects that correspond directly to entities in the real world or problem domain. This follows from the notion of domain modeling. As Grady Booch \cite{booch1991ooaod} puts it: ``Object-oriented analysis is a method of analysis that examines requirements from the perspective of the classes and objects found in the vocabulary of the problem.''

In combination, these two ideas imply that each significant computation of the software should be assigned to a single object that represents an entity in the problem domain. To be fair to advocates of object orientation, this implication is typically mitigated with an escape hatch allowing some computations to be assigned to objects that are \textit{not} problem entities. These objects are sometimes called `design objects' or `implementation objects' in contrast to the `analysis objects' or `domain objects' of the real world.

This mitigation does not solve the problem however, because it turns out that almost \textit{no} computation can be satisfactorily assigned to a domain object. Because a compelling assignment rarely exists, the programmer will often choose an arbitrary one. As a result, the legibility and modularity of the code suffers.

\section{Modeling Behavior}

To explain this problem properly, we need to construct an independent account of what computation might mean, viewed in the context of the problem domain. We'll do this using a restaurant reservation application as an example.

Describing behavior has three parts. First, we identify the \textit{phenomena}, which provide the vocabulary for talking about behavior. Second, we can characterize the possible \textit{behaviors} by constraining which phenomena can occur (and how they causally affect each other). Third, we can \textit{organize} the behaviors; typically this is done by grouping phenomena in some way, and in practice therefore precedes the second step.

Phenomena come in three kinds:
\begin{itemize}
\item \textbf{Individuals}. The individuals are the things that participate in the behavior. Usually these are called objects or entities, but the term \textit{individual} is preferable because it emphasizes the key property (which is to be distinct from other individuals and to have a persistent identity) without implying other notions.
\item \textbf{Relationships}. Relationships are facts that associate individuals with each other. They can also associate individuals with simple values. All observable properties of individuals are captured in these relationships; individuals don't have additional attributes or qualities that are distinct from the relationships they participate in.
\item \textbf{Actions}. Actions are what happens in the world. They involve individuals and change relationships. Actions are assumed to be atomic, meaning that they happen instantaneously without any apparent duration. To describe an activity that spans some time one defines actions that correspond to its start and end.
\end{itemize}

\textbf{Example: Restaurant reservations}. To describe the behaviors associated with restaurant reservations, we might have these phenomena:
\begin{itemize}
\item \textbf{Individuals}: \texttt{Alice} and \texttt{Bob} (who are users); \texttt{Maido} and \texttt{Rosetta} (which are restaurants); \texttt{Slot\_1}, \texttt{Slot\_2} and \texttt{Slot\_3} (which are slots available for reservation); \texttt{Reservation\_1} and \texttt{Reservation\_2} (which are reservations).
\item \textbf{Relationships}: \texttt{for (Reservation\_1, Slot\_3)} is the relationship that says that \texttt{Reservation\_1} is a reservation for \texttt{Slot\_3}; \texttt{by (Reservation\_1, Alice)} is the relationship that \texttt{Reservation\_1} is a reservation by (that is, belonging to) \texttt{Alice}; \texttt{time (Slot\_3, 7:30pm)}, \texttt{date (Slot\_3, January-10-2026)} and \texttt{restaurant (Slot\_3, Maido)} are the relationships that say that \texttt{Slot\_3} is for \texttt{7:30pm}, on \texttt{January 10, 2026} at \texttt{Maido}.
\item \textbf{Actions}: \texttt{reserve (Alice, Slot\_3):(Reservation\_1)} is the action in which \texttt{Alice} reserves \texttt{Slot\_3} resulting in reservation \texttt{Reservation\_1}; \texttt{cancel (Reservation\_1)} is the action in which \texttt{Alice} cancels \texttt{Reservation\_1}. 
\end{itemize}

Some phenomena are directly observable in the physical world. Others belong to the problem domain and are observable indirectly, and might not have existed were it not for the software. The users and restaurants, for example, are observable in the world; the slots and reservations exist only in the problem domain. Relationships likewise might belong to the physical world or to the domain. For the purposes of this discussion, we'll assume that all relationships are stored within the software system and all actions are performed at the interface between the software and the world, an assumption that Michael Jackson has shown must be discarded for a proper understanding of software requirements, especially for cyberphysical systems \cite{jackson1995worldmachine}.

A \textit{trace} is a sequence of actions, a story of what happened. At each point in the trace, from before the first action until after the last, there's a set of relationships that hold. Together the trace and its relationships comprise a \textit{behavior}. 

\textbf{Example: A partial behavior}. Here is a trace along with just some of the relationships that hold after one of the actions (and is therefore only part of a behavior):
\begin{verbatim}
createSlot (Maido, 6:00pm, January-10-2026): (Slot_1)
createSlot (Maido, 6:00pm, January-10-2026): (Slot_2)
createSlot (Maido, 7:30pm, January-10-2026): (Slot_3)
reserve (Alice, Slot_3): (Reservation_1)
  by (Reservation_1, Alice)
  for (Reservation_1, Slot_3)
  restaurant (Slot_3, Maido)
  time (Slot_3, January-10-2026)
  date (Slot_3, 7:30pm)
reserve (Bob, Slot_2): (Reservation_2)
cancel (Reservation_2)
redeem (Reservation_1)
\end{verbatim}

The story this trace tells is that three slots were created for Maido; Alice then reserved one of those slots; Bob reserved another slot, and then canceled it; Alice then turned up and redeemed her reservation (and was seated). The relationships after the \texttt{reserve} action tell us that the resulting reservation was by Alice, and for a slot at a given restaurant on a given day and time. Not even all the relationships that hold at that point are included; for example, \texttt{time (Slot\_2, 6:00pm)} holds too because even though the slot has not been reserved it exists with its start time.

To define the set of all valid behaviors, the standard method is to specify the relationships that hold initially, and then for each type of action to specify a precondition (which relationships must hold for the action to occur) and a postcondition (which relationships will hold after the action occurs). Despite its name, the postcondition mutually constrains the relationships before, the relationships after and the values of the action's arguments and results. The set of traces, and the relationships that hold at each point, follow from the action specifications by induction.

\textbf{Example: Specifying an action}. Here is an informal specification of the \texttt{reserve} action:
\begin{verbatim}
reserve (u: User, s: Slot): (r: Reservation)
	requires no reservation exists for s
	ensures a new reservation r is created that is for s and by u
\end{verbatim}

Put more formally, the precondition says that there is no relationship before the action happens of the form \texttt{for (\_, s)} where \texttt{\_} is any reservation and \texttt{s} is the particular slot being reserved here.  The postcondition says that  the following relationships will hold after:

\begin{verbatim}
for (r, s)
by (r, u)
\end{verbatim}

\noindent where \texttt{r} is the particular reservation that has been created, and \texttt{s} and \texttt{u} are the given arguments. The postcondition's description of the reservation as new actually implies an additional precondition: that no relationship of the form \texttt{for (r, \_)} holds.

Notice that the action specification is cast in general terms over \textit{types} of phenomena. Thus the specification itself defines the action of type \texttt{reserve} that comprises all its particular occurrences; the types \texttt{User}, \texttt{Slot} and \texttt{Reservation} appearing in the action signature represent the sets of individuals corresponding to all users, slots and reservations; and the words \texttt{for} and \texttt{by} in the precondition and postcondition refer to relations that contain relationships amongst particular individuals.

\section{Object-oriented Modularization}

For a large system, one would like to organize the description of the behavior into modules. Ideally, these modules would separate concerns \cite{dijkstra1974separation} and provide a suitable basis for a division of labor. Ideally, the module structure is carried through to the implementation.

Using our model of behavior, we can now represent this in a simple way. A modularization is just a set of (named) modules, with each (type of) phenomenon---individual, relation and action---assigned to one or more modules. Note that this is not necessarily a partition; a phenomenon might appear in more than one module. The assignment must be closed in the sense that when any action is included in a module, the individuals and relations that appear in its specification must be included too.

We can now make more precise the key principle of object-oriented modularization. Recall that this principle says that computations are assigned to objects and that the objects are the objects of the problem domain. In terms of our phenomena, this means:
\begin{itemize}
\item Each individual (type) will correspond to an object (class);
\item There is one module associated with each object (class).
\end{itemize}
We are interpreting the `operations' of Meyer's Single Target principle to be the actions of the behavior. The instance variables of objects will be the relations that the object participates in. The closure principle (that an action cannot be assigned to a module without the relationships it accesses) corresponds to the object-oriented principle that an object's methods should only access that object's instance variables.

There is an elegant parsimony in this scheme. No special effort is required to define the set of modules; nor indeed is a separate notion of module even required. The individual types will play dual roles: as sets of individuals and as modules governing their behavior.

Since the closure principle will make relations follow the actions that mention them, the the task of object-oriented modularization can be easily formulated. It is just to assign each action (type) to an individual (type). As we shall see, however, this turns out to be much harder than one might imagine.

\section{Assigning Actions to Individuals}

The first obvious challenge in assigning actions to individuals is that some actions involve multiple individual types. Take the \texttt{reserve} action for example, in which a user makes a reservation for a slot:
\begin{verbatim}
reserve (User, Slot): (Reservation)
\end{verbatim}
Which of the three individual types---\texttt{User}, \texttt{Slot} and \texttt{Reservation}---should this action be assigned to? (Or to put it in the more conventional object-oriented terms, which class should the \texttt{reserve} method be declared in?).

There's no good answer to this question. We might ask: which individual is the action \textit{by}? That would suggest the action should be assigned to \texttt{User}. But then almost every action will go the same way, and our modularization will collapse. Another approach is to follow Meyer's \textit{object motto}: ``Ask not first what the system does. Ask what it does it to!'' \cite{meyer1988oosc}. In that case, we might choose \texttt{Slot}, since arguably it's the slot that has been reserved. Although all these options seem equally plausible (or implausible), none is in fact correct. If the assignment is to be to any single individual, it must be to \texttt{Reservation}, since this action constructs a reservation, and only the constructor of a class can allocate its instances.

We might hope that the cancel action would be simpler to handle since it involves exactly one individual:
\begin{verbatim}
cancel (Reservation)
\end{verbatim}
If the effect of cancellation is to mark the reservation as canceled, then indeed \texttt{Reservation} seems like a good home for it. But suppose that canceled reservations are simply discarded, as if they had never existed. Given the low costs of storage and the value of retaining a full log of activities, this might not be the best design choice, but it's not an unreasonable one. In this case, although to a novice it might seem correct to assign the action to \texttt{Reservation}, anyone with a modicum of object-oriented programming experience will know that this cannot be done. In a modern object-oriented language, a class cannot deallocate one of its own objects. Instead, the object to be deallocated must be disconnected from one or more other objects that point to it. So this might suggest assigning the action to \texttt{Slot} along with the \texttt{for} relation that associates slots and reservations, or to \texttt{User} along with the \texttt{by} relation that associates users and reservations. Either way, it seems strange that an action about a single individual must actually be assigned to a different individual.

Returning to the \texttt{reserve} action, another troubling complication emerges. Suppose that, in addition to imposing the precondition that the slot not be previously reserved, we require that the user not already hold another reservation for a different slot at the same time and restaurant. If the action is to be colocated with the relevant state, we need to determine where the state will be stored. A common object-oriented move would be to assign the state to \texttt{Restaurant}, perhaps on the grounds that the restaurant is the scope in which the double booking is being checked. The closure rule would then require, awkwardly, that the \texttt{reserve} action belong to \texttt{Restaurant} too, despite having no restaurant as an argument (the restaurant being implicit, by relationship to the slot being reserved). 

What's happening here is that \texttt{Restaurant} is acting as a container for a collection of reservations. This is not a wise design decision, because it's vulnerable to likely changes in scoping. Suppose we decide that a user should not be able to book two reservations at the same time, even at different restaurants---a constraint that is indeed imposed by most reservation systems. In that case, the state will have to be moved out of \texttt{Restaurant}. But to where? To \texttt{User} perhaps?

The inescapable conclusion is that assigning state to individuals is a fool's errand, and that the very premise of object orientation---that state and actions naturally belong to individuals---is mistaken.

\section{Impacts on Modularity}

We've seen that a core design principle of object orientation---namely that computations (actions and their supporting state) should be assigned to objects---is far harder to achieve than is normally recognized. If not arbitrary, the choice of object to which the computation is to be assigned can only be reliably evaluated by considering detailed properties of the code. The principle is thus hard to apply at all in the early stages of development, casting some doubt on the entire enterprise of object-oriented analysis and design. But even if one is able to assign computation to objects, the result is often poor modularity.

There are two key failures of modularity in software: \textit{fragmentation} (in which functions that belong together are fragmented amongst modules) and \textit{conflation} (in which functions that should be separated into distinct modules are conflated in a single module). These lead to additional problems: fragmentation leads to coupling (since the fragments need to be coordinated across modules) and conflation prevents reuse (since it results in modules that embody application-specific combinations of functions).

Here are some illustrations from our restaurant reservation system. As we saw earlier, in order to check that a reservation has not already been made at a given time, the \texttt{reserve} action and the relationships that map reservations to users and times might be assigned to the \texttt{User} object. But presumably the actions and state associated with user authentication will be assigned to the same object, along with other functionality involving users (such as karma tracking, communication preferences for notifications, restaurant reviews and ratings, and so on). This is conflation, with many unrelated functionality concerns entangled in a single object. Since that object combines a collection of concerns that correspond to the system's features, and no two systems would be likely to have exactly the same set of concerns, the \texttt{User} object becomes almost a fingerprint for the system as a whole, reflecting the extent to which reuse has been compromised.

Considering the \texttt{reserve} action again, recall that there are two invariants we would like to maintain: (a) that there is only one reservation per slot, and (b) that a user has only one reservation at a given time. The first invariant can be ensured by assigning the action to \texttt{Slot}; the second by assigning it to \texttt{User}. To maintain both invariants, we will have to abandon the single assignment rule, and assign the action to both objects. And as noted earlier, since the action requires the creation of a new reservation object, it must be assigned to \texttt{Reservation} too. The net result will be three methods, one in each of the classes, along with associated relationships in each (from slots to reservations in \texttt{Slot}, from users to reservations in \texttt{User}, and from reservations to restaurants in \texttt{Reservation}). The functionality will thus be fragmented across three objects, and in particular the rules that prevent overlapping reservations, embodied in the two invariants, will be split across objects too. This fragmentation leads also to coupling, because executing the \texttt{reserve} action will require coordinating the three methods that the action corresponds to in the three objects.

\section{The problem of navigation}

A different facet of object orientation leads to another modularity problem: that the direction of intended navigation often determines the assignment of relationships to objects, resulting in undesirable couplings or dependencies \cite[pp. 275-276]{jackson2021essence}.

Suppose our restaurant system allows users to submit reviews of restaurants. To model this behavior, we might add a new individual type \texttt{Review}, a new action

\begin{verbatim}
review (User, Restaurant, Text): (Review)
\end{verbatim}

\noindent and new relations that associate reviews with their authors, target restaurants and textual content. Now consider the relation that associates reviews and restaurants. Each review is for exactly one restaurant, and restaurants will have multiple reviews. This suggests that the most natural assignment of the relation is to \texttt{Review}. But because at runtime there will be need to display a restaurant's reviews along with the restaurant details, an object-oriented programmer will be inclined instead to assign the relation to \texttt{Restaurant}, represented in the code as an instance variable holding a set of reviews, and the action is likely to follow as a method of \texttt{Restaurant} (even as it will also require a constructor in \texttt{Review}). As a result, there will be a dependency in the code of \texttt{Restaurant} on \texttt{Review}, with \texttt{Restaurant} not only mentioning the \texttt{Review} class in its code but likely calling its constructor too.

In his seminal paper on dependencies \cite{parnas1979extension}, David Parnas presents a compelling rule for when one module should depend on another:

\begin{em}
We propose to allow A `uses' B when all of the following conditions hold:
\begin{itemize}
\item A is essentially simpler because it uses B;
\item B is not substantially more complex because it is not allowed to use A;
\item there is a useful subset containing B and not A;
\item there is no conceivably useful subset containing A but not B.
\end{itemize}
\end{em}

Let us apply this rule to our case, in which \texttt{Restaurant} depends on \texttt{Review}. The first two criteria do not seem to hold. The third criterion doesn't seem to hold either; a subset containing \texttt{B} but not \texttt{A} would be a version of the restaurant reservation system that has reviews but no restaurants; that might be the basis for a different kind of system, but it certainly wouldn't be a restaurant reservation system. The fourth criterion is the most clearly violated, since it says that there should no subset of the restaurant system with restaurants but no reviews, which is obviously wrong.

This example is not an outlier but rather represents a common pattern in object-oriented programming. In short, the most natural navigation patterns produce dependencies in exactly the \textit{opposite} direction from that which a consideration of modularity would suggest.

\section{Prior Mitigations}

None of these problems are new; all have been discussed in research papers over the last several decades, and a variety of embellishments and workarounds have evolved to mitigate them. 

To address the problem that actions often cannot be naturally assigned, new categories of objects are usually added that do not correspond to problem domain individuals. Meyer, for example, proposes three categories of classes---analysis classes, design classes and implementation classes---only the first of which corresponds to individuals in the problem domain \cite{meyer1988oosc}. When the scope of an action needs to incorporate multiple objects, a `collection' object is often included; thus in addition to \texttt{Reservation}, there might be a class called \texttt{Reservations} each instance of which holds a set of reservations.

Domain-driven design \cite{evans2003ddd} recognizes both the need for collection classes and the problem of fragmentation of functionality across objects, and introduces \textit{aggregates}, which are objects that are intended to contain all the objects associated with a given function and to preserve cross-object invariants. For example, a \texttt{Reservations} aggregate may contain all \texttt{Reservation} and \texttt{Slot} objects.

Many strategies have been proposed to address conflation. The earliest are multiple inheritance and mixins, but due to the complications they bring, they have not been widely adopted and are not included in recent object-oriented languages. Subject-oriented programming \cite{ossher1993subject}, and later hyperslices \cite{tarr1999ndegrees}, allowed a single object to have different fields for different purposes, but were perhaps too disruptive to standard object-oriented patterns to be widely adopted. The entity-component system \cite{bilas2002datadriven} achieves a similar effect through more traditional means, by representing the different projections of an object as distinct subobjects that an object is explicitly mapped to (by a global hash table, for example) and to which methods can be delegated. Aspect-oriented programming \cite{kiczales1997aop} also addressed the difficulty of separating concerns in an object-oriented program but was more focused on factoring out application-wide concerns that `cross cut' other concerns than on undoing conflation within individual objects. Its solution is to provide additional layers of functionality with their own objects, whose methods are invoked by hooks that intercept method calls in a base layer.

Role-oriented programming \cite{reenskaug1996ooram}, which was later realized more fully in the DCI framework \cite{coplien2009dci}, has some motivations in common with these other approaches, but its primary goal is to make code more readable, by having objects and methods correspond more closely to the individuals and actions of the domain. Use cases are implemented as \textit{contexts} in which the underlying objects are given distinct \textit{roles}; thus a slot object might have a role for the interaction in which its properties are established, and a different role for the interaction in which it is assigned to a reservation. Roles are objects in their own right that delegate to underlying problem domain objects. Conflation and fragmentation are addressed by expressing as much functionality as possible in the role objects, and the domain objects are treated almost as a low-level database.

The problems caused by navigation have also been widely recognized. The most ambitious proposals have involved augmenting object-oriented languages with richer ways to handle relationships. RelJ, for example, lets programmers define relationships independently of objects, in their own type hierarchy \cite{bierman2005firstclass}.

The troublesome dependencies that arise as an inevitable byproduct of object orientation led to design patterns that seek to mitigate the dependencies by introducing abstract interfaces and indirections \cite{gamma1994designpatterns}. Although often elegant and effective in eliminating dependencies on particular classes, the net effect of these patterns is to make code far harder to understand. By introducing new computational steps (to maintain the pattern structures), they weaken the link between the computation apparent in the code and actions in the problem domain. As the design patterns book explains: 

\begin{quote}
An object-oriented program's run-time structure often bears little resemblance to its code structure. The code structure is frozen at compile-time; it consists of classes in fixed inheritance relationships. A program's run-time structure consists of rapidly changing networks of communicating objects. In fact, the two structures are largely independent. Trying to understand one from the other is like trying to understand the dynamism of living ecosystems from the static taxonomy of plants and animals, and vice versa.
\end{quote}

In one of this talks, Trygve Reenskaug called this `a frightening observation,' perceptively attributing to these patterns much of the difficulty we now have in reading or analyzing object-oriented code.

\section{The Wider Impact of Object Orientation}

One may wonder how widely the critique of this paper applies, given that many software systems today are not traditional object-oriented programs. Most web applications, for example, are built on top of relational databases which allow more flexible treatment of actions and relationships.

In practice, however, the influence of object orientation remains. All popular web frameworks include an `object-relational mapper' (or are compatible with third-party plugins). This allows queries and updates to a relational database to be expressed as method calls over a graph of objects that is constructed on the fly from the database's tables. The resulting code looks more familiar and may express database accesses more succinctly. But the imposition of an object-oriented structure brings all the problems that we have raised in this paper.

The influence of object-oriented thinking is so pervasive that many systems tend to follow object-oriented principles implicitly and to pay the price accordingly. This phenomenon is evident in RealWorld \cite{realworldbenchmark}, a benchmark web application for which several hundred instances are available in different languages and platforms. Almost all of the submissions are described by their authors as demonstrations of best practice, and the benchmark problem---a simplified microblogging app in the style of Medium called Conduit---is small enough to allow careful polishing. Looking at the code reveals the same problems of assigning functions to objects, albeit at a slightly coarser granularity.

A typical structure for the code breaks it down in two dimensions. There is a division into horizontal layers, each intended to encapsulate an aspect of functionality: typically a routing layer at the top (encapsulating the details of HTTP requests and responses, and the application-specific design of URLs); a controller layer in the middle (encapsulating the core logic of the application); and a model layer at the bottom (encapsulating the design of the data model and the choice of database). There is also commonly a division into distinct vertical stacks that has a strongly object-oriented flavor. For example, the stacks might be associated with the individuals \texttt{User}, \texttt{Article} and \texttt{Comment}, with a file corresponding to each layer and stack (thus, \texttt{user\_routing}, \texttt{article\_routing}, \texttt{comment\_routing}, \texttt{user\_controller}, \texttt{article\_controller}, \texttt{comment\_controller}, etc).

Consequently the assignment of functionality to these stacks suffers from the core problem of object orientation that we have discussed, leading to conflation and fragmentation. The fragmentation causes couplings across vertical stacks, despite the general intent of the design to keep the stacks independent of one another. For example, in one of the submissions to the RealWorld benchmark \cite{winterrrrrff} the action for favoriting an article is assigned to the controller layer of the \texttt{Article} stack. This function calls a function in the model layer of \texttt{Article} that updates the favorites count for the article. So far so good. Unfortunately, it also calls a function in the model layer of the \texttt{User} stack that updates a list of favorited articles associated with a user. And in fact the function setting the favorites count in the model layer of \texttt{Article} obtains that count from a newly updated count in the model layer of \texttt{User}.\footnote{Further evidence that the developer has struggled with assigning this favoriting action to a single stack can be seen in code in the \texttt{User} model function that actually calls a function in the \texttt{Article} stack---a dependency in the other direction---that has been commented out!}

\section{A Simpler Approach to Modularity}

As we noted above, the key design principle of object orientation is that domain objects play dual roles, not only as digital twins for individuals in the real world, but also as modules for the organization of functionality. This approach has a compelling elegance to it, and an attractive parsimony in not requiring any additional notion of module. But as we have seen, the assignment of functionality to objects is far more complicated than it might appear to be at first sight. The assignments that result tend to be arbitrary, and lead both to fragmentation and conflation.

A simpler approach is possible that offers stronger modularity properties. Rather than assigning function to the individuals of the problem domain, we introduce a new category of modules which we call \textit{concepts} \cite{jackson2021essence}. Each concept embodies some actions and some relations over the individuals that participate in those actions. While the concepts of a system partition the actions and relations of the behavior, they do \textit{not} partition the individuals. A single individual can participate in actions, and be mapped by relations, in multiple concepts.

Two particular features of concepts enable stronger modularity. First, since an individual can be mapped by relations in different concepts, an object can essentially be split into distinct views. For example, a \texttt{UserAuthentication} concept might map individuals of the type \texttt{User} to passwords, while the concept \texttt{UserProfiling} maps the same individuals to display names and bios. This allows concepts to represent separate concerns, preventing conflation of functionality.

Second, concepts never call each other. There is no delegation with one concept `using' another concept; each concept stands alone and delivers its own functionality (of course making use, in the implementation, of underlying platform libraries). Instead, concept behaviors are coordinated by \textit{synchronizing} their actions, with declarative rules that establish causal links between them \cite{mengjackson}.

Despite these differences, the most fundamental idea of object orientation remains: that individuals in the problem domain are reified in the software as objects with persistent identity. Actions and state are not partitioned amongst those objects, however, but are assigned to larger modules that encompass collections of objects. Because a concept can store sets of objects, it is no longer necessary to introduce new object types to represent collections. 

The concept approach can be straightforwardly implemented in any programming language. In an object-oriented language, it is appropriate to represent each concept as a class that is instantiated to obtain an instance with its own state. A concept is thus an object in the sense of the tiny machines view of object orientation. In our account so far, there is no need for more than one instance of a given concept, but multiple instances are possible (and indeed may be desirable to support better resource allocation, using sharding for example).

Although the concept approach differs from traditional object-oriented programming, it aligns with several contemporary trends in software development. Concepts are similar to microservices, although more fine grained. There is no requirement that concepts be separately deployed, although that is straightforward, and in that respect a basic concept-structured app is similar to what has been called a modular monolith. Action synchronization is very similar to event-driven architectures, with the synchronization rules playing a similar role to the anti-corruption layer of domain driven design. An important difference is that concepts aim to provide legibility in the code by ensuring a clear correspondence between code elements and problem domain features. So, unlike event-driven architectures in which API functions publish and subscribe to events, there is only one category of actions that play the role of API functions, bus events and domain events all at once.

\section{An Example}

\begin{figure}
{\small
\begin{verbatim}
concept Reserving [ User, Slot ]

purpose manage commitments to users for subsequent access to resources

principle
  after a user reserves a slot for a given date and time and party size,
  then if they turn up at that date and time with a party of that size
  they will be granted access to the slot

state
  a set of Reservations with
    a User
    a Slot
    a Time
    a Date
    a partySize Number
    a canceled Flag
    a granted Flag

actions
  reserve (user: User, slot: Slot, time: Time, date: Date, partySize: Number)
    : (reservation: Reservation)
    requires no existing reservation for slot,
      or for user on date and at a time close to time
    ensures creates a new reservation with canceled and granted false

  cancel (reservation: Reservation)
    requires reservation exists and is not canceled or granted
    ensures marks r as canceled

  grant (reservation: Reservation)
    requires reservation exists and is not canceled,
      and the current time and date is close to reservation's time and date
    ensures marks reservation as granted
    
  noShow (reservation: Reservation)
    requires reservation exists and is not canceled or granted,
      and the current time and date is after reservation's time and date,
    ensures marks reservation as canceled
\end{verbatim}
}
\caption{The Reserving concept}\label{reserving-fig}
\end{figure}

\begin{figure}
{\small
\begin{verbatim}
concept Availability [ Venue ]

purpose manage availability of resource slots at different venues

principle
  after some slots have been created for a venue at different times,
  one can get the available slots, and then toggle the availability of a slot
  to indicate that it has been assigned externally

state
  a set of Slots with
    a Venue
    a start Time
    an end Time
    a Date
    a minParty Number
    a maxParty Number
    an available Flag

actions
  create (venue: Venue, start: Time, end: Time, date: Date, max, min: Number): (slot: Slot)
    requires end is after start, max >= min
    ensures creates a new available slot with given properties
  
  toggleAvailable (slot: Slot)
    ensures toggles the available flag for slot
  
  _getAvailableSlot (time: Time, date: Date, partySize: Number): (slot: Slot)
    returns any available slot with start time matching time,
    and date matching given date and min <= partySize <= max
\end{verbatim}
}
\caption{The Availability concept}\label{availability-fig}
\end{figure}

\begin{figure}
{\small
\begin{verbatim}
sync makeReservation
when Requesting.reserve (restaurant, time, date, partySize, session)
where
  Availability._getAvailableSlot (time, date, partySize) : (slot)
  UserAuthentication._getUser (session) : (user)
  Karma._inGoodStanding (user)
then
  Reserving.reserve (user, slot, time, date, partySize)

sync punishNoShow
when Reserving.noShow (reservation)
where reservation.user = user in Reserving
then Karma.punish (user)
\end{verbatim}
}
\caption{Sample synchronizations}\label{syncs-fig}
\end{figure}

The figures shows examples of concepts (Figs. \ref{reserving-fig}  and \ref{availability-fig}) and synchronizations (Fig. \ref{syncs-fig}) for a simplified restaurant reservation system. There are two sample concepts, \texttt{Reserving} and \texttt{Availability}. Some additional concepts are implied by the synchronizations but not shown here: a \texttt{UserAuthentication} concept (which generates session tokens and provides a query that given a session token returns the associated user); a \texttt{Karma} concept that tracks good and bad behavior of users in the standard way; and some concept, say \texttt{RestaurantCatalog}, that generates restaurant individuals (and that supports finding restaurants and viewing their information).

\texttt{Reserving} and \texttt{Availability} provide subtly different functionality. \texttt{Reserving} is about making commitments to users that certain resources will be available at a particular later time; \texttt{Availability} is about predicting that availability based on an allocation of resources. But neither the prediction nor the commitment are certain; there may be more or fewer tables available than the slot allocation suggested, and it may in fact not be possible to fulfill the commitment of a reservation. The actual allocation will be governed by another concept (that assigns reservations to tables). The slots that are predicted by the \texttt{Availability} concept are in practice not mapped simply to tables, both because tables are often reconfigured on the fly to accommodate parties of different sizes, and because of other complexities (such as overbooking).\footnote{These considerations explain why an expected invariant does not hold. One might have assumed that the time of a reservation in the \texttt{Reserving} concept matches the start time of its associated slot in the \texttt{Availability} concept, but in practice these can diverge (for example if a diner calls to say they will be late) and have different meanings: the first reflects a commitment to a customer, and the second is a prediction of availability.}

There is one special concept that appears in a synchronization that merits some explanation. Sometimes, a user's request to perform an action is rejected, for example because the user does not have the requisite permission. In an earlier semantics for synchronizations \cite{jackson2021essence}, this was handled by treating a synchronization as a symmetric constraint on all its actions, following CSP \cite{hoare1985csp}. The sync would specify that the action requested was accompanied by the action of permission being granted; if either failed, neither could happen. In practice, designers find this approach hard to understand: intuitively, a request for an action occurs, and the action can fail, but in this approach the action and its request are inseparable and if the action fails, it is as if the request never occurred. It is also hard to implement, since it requires transactions, and provides no simple way to log unsuccessful requests.

The approach \cite{mengjackson} used here is much simpler. Synchronizations are rules that represent causal links: when one action happens, another might follow. In this approach, requests are reified as actions in their own right, so that a request can occur without the action necessarily following. This works well in practice, because it also allows the request and the action to have different arguments; a user's request may include only a session token, for example, with a \texttt{UserAuthentication} concept providing a query to translate it into a user identity. We typically put all user requests in a single pseudo-concept called \texttt{Requesting}; in the implementation, this concept encapsulates the details of HTTP and routing.

Each concept contains the following parts:
\begin{itemize}
\item \textbf{Name}. The name of the concept can be followed by a list of type names, which are treated as type variables in the body of the concept. Thus \texttt{concept Availability [ Venue ]} says that the Availability concept is parameterized by a type whose individuals represent venues (which in the context of the system will be restaurant individuals).
\item \textbf{Purpose}. The purpose expresses the motivation for including the concept's functionality. The purpose is as specific as possible, in order to most clearly identify the concern that this particular concept provides. 
\item \textbf{Principle}. The operational principle is an archetypal scenario that illustrates how the behavior of the concept fulfills the concept's purpose.
\item \textbf{State}. The state section declares the relations (amongst individuals, and between individuals and simple values). The relations are organized for succinctness around individual types; this does not imply any kind of object structuring. Thus, for example, \texttt{a set of Reservations with a User} introduces a relation from reservations to their users. For convenience, the relation names can be explicit (as in \texttt{a partySize Number}), or implicit, with the relation name being the lower-cased form of the type name (as in \texttt{a User}, which is equivalent to writing \texttt{a user User}).
\item \textbf{Actions}. The actions section specifies each of the concept's actions in standard precondition/postcondition form. The preconditions are to be interpreted as firing conditions: that is, if the precondition does not hold, the action may not occur. Result arguments are named (so multiple results can be returned). The section can also specify queries, which are marked by an initial underscore. A query is not an action, and cannot update the state, but is just a convenient way to specify a common state-reading pattern. A query always returns a set of bindings; thus \texttt{\_getAvailableSlot} returns a set of slots (which will be either a singleton with an available slot, or an empty set if no such slot exists). The state of a concept is assumed to be public, so there is no need to define a rich set of queries akin to the observers of an abstract data type.
\end{itemize}

Several details of timing are left underspecified---for example, how much have time must have passed after a reservation to regard it as a no show or to no longer grant the reservation (seating the party). Trickier is the question of how close in time two distinct reservations can be to allow a user to hold both. Popular reservation systems may not allow two reservations on the same night, but customers complain about this (because it prevents them from booking one restaurant for cocktails first and another for a meal after).

A synchronization has four parts:
\begin{itemize}
\item \textbf{Name}. The name of the synchronization is used in provenance tracking to point to the causal link that resulted in one action following another.
\item \textbf{When}. The \texttt{when} clause provides a pattern for the action occurrences that fire the synchronization. For example, the clause in the first sync mentioning \texttt{Requesting.reserve} says that the sync fires when a request for a reservation occurs, and the clause in the second sync mentioning \texttt{Reserving.noShow} says that the sync fires when a \texttt{noShow} action occurs in the \texttt{Reserving} concept. The variables that appear in the pattern are bound by name to the action arguments (which allows partial matching, not illustrated here, in which not all arguments are included).
\item \textbf{Where}. The \texttt{where} clause binds additional variables by querying the state of one or more concepts. For example, in the second sync, the \texttt{where} clause obtains (from the \texttt{Reserving} concept's state) the user associated with a no-show reservation. In the first sync, the clause does more work: it obtains the user associated with the session token passed in the request; finds a matching available slot for the given restaurant, date, time and party size; and checks that the user is in good standing (that is, hasn't been punished too many times for no shows).
\item \textbf{Then}. The \texttt{then} clause specifies an action that is to be executed. In general, the specified action is executed for every possible binding that the \texttt{where} clause produces\footnote{This allows syncs to express iterations easily. For example, a sync might specify that \textit{when} a user is deleted, \textit{where} a reservation belongs to the user, \textit{then} the reservation is deleted. If a user has multiple reservations, the reservation deletion action will then be performed once for each such reservation.}. For the first sync, there is at most one such binding (and no binding if the session token is invalid, so no user is returned by the \texttt{UserAuthentication} query, or if the user is not in good standing according to the \texttt{Karma} query). For the second sync, there is always one binding, with each no show resulting in the user being punished.
\end{itemize}

This example illustrates some key respects in which concept and object-oriented structures differ:
\begin{itemize}
\item \textbf{No composite objects}. The state is expressed as a collection of relations amongst individuals (and values), and not as composite objects. Individuals are passed between concepts by synchronizations, but since they have no composite structure, none of the problems of sharing mutable objects arise. 
\item \textbf{No methods}. Actions do not have a single target, but can have any number of arguments that play an equal role, and their behavior can be defined over the entire concept state.
\item \textbf{No non-domain types}. Since the concept state comprises sets of individuals and their relationships, there is no need for collection classes or other kinds of non-domain object types. Of course, in an implementation there will be additional types (for the bindings that arise in synchronizations, for example) but these are not needed to describe behavior in the analysis and design phases. This alignment with the types of the problem domain contributes to the legibility of the specifications and code.
\item \textbf{Separation of concerns}. Concepts allow a fuller separation of concerns, because the same individual can be involved in different relations and actions in different concepts. Slots, for example, play different roles in the \texttt{Reserving} and \texttt{Availability} concepts. In the former, they are associated with reservations and represent the resources being reserved; in the latter, they are associated with minimum and maximum party sizes and represent the resources that are available. In this example, it happens that \texttt{Slot} sits on the right hand side of the relation in \texttt{Reserving}, but that is not necessary. The concept could equally well declare \texttt{a set of Slots with a Reservation}, so that both concepts declare sets of slots with their concept-specific properties.
\item \textbf{Externalized coordination}. There are no calls from one concept to another, nor indeed any kind of reference within a concept to another concept. All coordination between concept is expressed in synchronizations. This ensures that concepts are free standing and reusable, and are less polluted with application-specific details (for example, that no shows result in a particular degree of karma punishment).  Moreover, externalizing coordination in this way makes it much easier to track the causal links between actions.
\item \textbf{Polymorphism}. Concepts are inherently unboundedly polymorphic. Thus the \texttt{Reserving} concept, for example, does not depend on users or slots having particular types, but requires only that one user or slot be distinguishable from another.
\end{itemize}

\section{Conclusion}

Concept structuring seems to address the limitations of object orientation, offering a clearer correspondence between problem domain and design elements, and a more effective separation of concerns. Experience in ongoing industrial application has been positive, where concepts also play a vital role as a shared language between team members with different specialties (in particular bridging between engineers and user experience architects). At the implementation level, the independence of concepts (and their avoidance of fragmentation and conflation) makes them well suited to LLM-based code generation \cite{mengjackson} \cite{jacksonWYSIWYD}, since both concept specifications and implementations can be generated one at a time, without having to include other concepts in the context.

Of course challenges remain. Although concepts are simple, they're not easy. Programmers, and even software designers and architects, are psychologically drawn towards mechanism and away from abstraction, and disentangling concerns is hard (if rewarding) work.

Some aspects of behavior, especially those in which the composite structure of objects is relevant, are well matched to an object-oriented perspective, and these can be hard to capture with concepts. For example, creating fingerprints or hashes of objects, serializing them, or describing versioning, all involve computing a function on object contents defined by some object boundary. These computations can be defined in a concept setting, but are harder to generalize since they must be parameterized over the relations that comprise an object's contents.

The essential idea of concepts might be called \textit{relational modularity}. Approaches that were developed to improve the modularity of object-oriented programming, notably aspect-oriented programming \cite{kiczales1997aop} and the bounded context idea of domain-driven design \cite{evans2003ddd}, sought modularity by scoping \textit{objects}. Concepts instead scope \textit{relations} and their \textit{invariants}. 

All the core elements of this approach were present in the 1990s, in datalog and deductive databases; in formal specification languages such as Z and B (and especially in the work on view specification \cite{jackson1995structuring,ainsworth1994viewpoint,derrick1995crossview,finkelstein1992viewpoints}); in semantic data models such as the entity-relationship model \cite{chen1976er} and semantic networks \cite{LEHMANN19921}. But the organizational structure of systems, prior to the emergence of domain-driven design and microservices, was dominated by global schemas and monolithic architectures. The pressing problems back then were storage efficiency, transaction correctness and query optimization, and the problems of semantic drift, model evolution and organizational alignment were yet to be regarded as important.

Today software development looks very different. Fast and capacious hardware makes a semantics-driven approach possible. Team ownership of models and bounded contexts are widely accepted; event-driven architectures are ubiquitous. Given all these changes, perhaps the time has come to reconsider a core assumption that has shaped programming for decades, and to explore new forms of modularity that will serve software developers better in the future.

%
%
%
\bibliographystyle{splncs04}
\bibliography{refs} 
%
\end{document}